\documentclass [12pt] {article}

\usepackage[square,sort&compress,comma,numbers]{natbib}

\usepackage{amsbsy}
\usepackage{amsmath}
\usepackage{graphicx}
\usepackage{graphics}
\usepackage{multirow}
\usepackage{epsfig}
\usepackage{setspace}
\usepackage{color} 
\usepackage{colortbl} 
\usepackage{stackrel}

\usepackage{verbatim}
\usepackage{float}
\usepackage[margin=1in]{geometry}
\usepackage{natbib}
\usepackage{epstopdf}

\usepackage{authblk}
\usepackage{hyperref,lineno}
\usepackage{siunitx}
\usepackage{upgreek}
\usepackage[font=small, labelfont = bf]{subcaption}
\usepackage[font=small,labelfont=bf]{caption}
\usepackage[dvipsnames]{xcolor}
\usepackage{listings}
\usepackage{xcolor,soul}
\soulregister\cite7
\usepackage{longtable}
\usepackage{booktabs}

\begin{document}

\title{A Bayesian regularization-backpropagation neural network model for peeling computations}
			
\author[1]{\small{Saipraneeth Gouravaraju}}
			\affil[1]{\footnotesize{Indian Institute of Technology Guwahati, Guwahati, India 781039}}
			
\author[1]{\small{Jyotindra Narayan}}
			
\author[1,2,3,4]{\small{Roger A. Sauer}}
			\affil[2]{\footnotesize{Aachen Institute for Advanced Study in Computational Engineering Science (AICES), RWTH Aachen University, Templergraben 55, 52056 Aachen, Germany}}
\affil[3]{Department of Mechanical Engineering, Indian Institute of Technology Kanpur, UP 208016, INDIA}
\affil[4]{ Faculty of Civil and Environmental Engineering, Gda\'{n}sk University of Technology, ul. Narutowicza 11/12, 80-233 Gda\'{n}sk, Poland}
			
\author[1]{\small{Sachin Singh Gautam}\footnote{Corresponding Author, email: \href{mailto:ssg@iitg.ac.in}{ssg@iitg.ac.in}}}
			
\date{}
			
\maketitle

\begin{abstract}
A Bayesian regularization-backpropagation neural network (BR-BPNN) model is employed to predict some aspects of the gecko spatula peeling, viz. the variation of the maximum normal and tangential pull-off forces and the resultant force angle at detachment with the peeling angle. $K$-fold cross validation is used to improve the effectiveness of the model.  The input data is taken from finite element (FE) peeling results. The neural network is trained with $75\%$ of the FE dataset. The remaining $25\%$ are utilized to predict the peeling behavior. The training performance is evaluated for every change in the number of hidden layer neurons to determine the optimal network structure. The relative error is calculated to draw a clear comparison between predicted and FE results. It is shown that the BR-BPNN model in conjunction with the $k$-fold technique has significant potential to estimate the peeling behavior.
\end{abstract}


\section{Introduction}\label{introduction}
The study of peeling is essential in understanding the adhesion characteristics in many applications such as adhesive tapes, micro- and nano-electronics \cite{Komvopoulos2003,Zhang2009}, coatings \cite{Sexsmith1994}, microfiber arrays \cite{Majidi2006,Schubert2007}, wearable medical bands \cite{Drotlef2017}, and cell adhesion \cite{Zhu2000}. Peeling problems have been used by many researchers to analyze multiscale adhesion in biological adhesive pads such as in geckos, insects, and spiders \cite{Persson2003,Sauer2009,Labonte2016,Federle2019}, where peeling is an important aspect of detachment.

Peeling, particularly gecko spatula peeling, has been studied extensively using experimental~\cite{Autumn2000,Autumn2002}, analytical\cite{Tian2006}, and computational methods~\cite{Sauer2011,Sauer2013,GautamIJNME2013,GautamIJCM2014,AgrawalSadhana2021}. However, each of these methods comes with its specific limitations. Although experimental methods provide insights into the gecko adhesive system, they are limited in resolution, typically at seta level. To the authors' best knowledge, there have been no experimental studies that explored the adhesive and frictional behaviour at the spatula level owing to the difficulty in isolating a single spatula. Most of the analytical models that study the peeling of gecko spatulae, although they provide insights into the various aspects of the peeling behaviour, they are limited by their inherent assumptions such as steady-state peeling, zero bending stiffness, and linear material response. As such, most of the analytical models are unable to predict the entire peel-off process, including the snap-off behaviour. This necessitates the use of a numerical analysis tool like FEM. However, the computational cost can  become very high due to the nonlinear and small scale nature of molecular adhesion as well as the detailed spatula microstructure.  The high computational cost can be overcome by reduced models, such as beam models~\cite{SauerMergelFEAD3DBeam2014}, but the cost remains a major limitation of full continuum models. Recently, Gouravaraju et al.~\cite{Gouravaraju2020a,Gouravaraju2020b} have studied the peeling behaviour of a single gecko spatula. However, as mentioned above, the computational cost of the numerical model is very high.  As observed by some authors ~\cite{Gu2018,Oishi2020,Kim2020}, the use of machine learning techniques such as artificial neural networks has the potential to reduce these computational costs while retaining the accuracy of numerical methods. In particular,  Gu et al.~\cite{Gu2018} have shown that employing neural networks can significantly reduce the high computational cost of FE simulations. To the best of the authors' knowledge there has been no study that employs machine learning techniques to analyze adhesive peeling and specifically gecko spatula peeling. Therefore, in this work, a Bayesian regularization-based backpropagation neural network~ \cite{Argatov2019,MacKay1992,Burden2008}  is employed to predict the influence of the peeling angle on the peeling force of a gecko spatula. The input data is obtained from the finite element simulations of Gouravaraju et al.~\cite{Gouravaraju2020a,Gouravaraju2020b}, who have used a quasi-continuum finite element model that captures friction due to adhesion at the nanoscale~\cite{Sauer2007,Mergel2019}.

The remainder of the paper is structured as follows: Section 2 discusses the adhesive friction model and the peeling of the spatula. In section 3 a backpropagation neural network with Bayesian regularization is presented. Section 4 discusses the implementation of the neural network model. Results and discussion are presented in section 5. Finally, section 6 concludes the paper.

\section{Peeling using an adhesive friction model}\label{peeling_explanation}
\label{sec:peeling}
In this section, the adhesive friction model of Mergel et al.~\cite{Mergel2019} and its application to gecko spatula\footnote{The nanoscale spatulae in geckos are very thin structures (approximately $5-10$ nm thick) with a width of around $200$~nm that can be modeled effectively as a thin strip \cite{Tian2006,Pesika2007,Peng2010,Sauer2011}.} peeling by Gouravaraju et al.~\cite{Gouravaraju2020a,Gouravaraju2020b} are briefly described.

The ``Model EA" of Mergel et al.~\cite{Mergel2019} defines a sliding traction threshold $T\!_\mathrm{s}$ that is non-zero even for tensile normal forces. This sliding threshold depends on the magnitude of the normal traction $T\!_\mathrm{n}=\Vert \boldsymbol{T}\!_\mathrm{n}\Vert$ due to adhesion between the spatula and the substrate. Further, it is assumed that the interfacial frictional forces act only up to a certain cut-off distance $r\!_\mathrm{c}$. Then we have,
\begin{equation}
T_\mathrm{s}(r) = \begin{cases}
\displaystyle\frac{\mu^{}_{\mathrm{f}}}{J_{\text{c}}}\Big[T\!_\mathrm{n}(r) - T\!_\mathrm{n}(r_\mathrm{c})\Big], &  \quad r < r_{\text{c}}, \\
\displaystyle \quad \quad \quad 0, & \quad r \geq r_{\text{c}},
\end{cases}
\label{eq:sl_thresh}
\end{equation}
where $J_\mathrm{c}$ is the local contact surface stretch ($=1$ for rigid substrates), $\mu_\mathrm{f}$ is the friction coefficient, and $r$ denotes the distance to the substrate surface.

The normal traction $\boldsymbol{T}\!_\mathrm{n}$ is obtained from the variation of the total adhesion potential, which is the summation of individual adhesion potentials acting between the molecules of the substrate and the spatula, and is given as~\cite{Sauer2009a}
\begin{equation}
\label{eq:Adhesive_Traction_Ref}
\boldsymbol{T}\!_\mathrm{n} = \frac{{A}}{2\pi r_0^3}\,\left[  \frac{1}{45} \left(\frac{r_0}{r}\right)^9 - \;\; \frac{1}{3} \left(\frac{r_0}{r}\right)^3  \right]\,\boldsymbol{n}_\mathrm{s}\,,
\end{equation}
where $r_0$ is the equilibrium distance of the Lennard-Jones potential,  $A$ is Hamaker's constant, and $\boldsymbol{n}_\mathrm{s}$ is the normal to the substrate.

Similar to Coulomb's friction model, the magnitude of frictional traction $\boldsymbol{T}\!_\mathrm{f}$ is governed by
\begin{equation}
\label{eq:friction_traction}
\left\Vert\boldsymbol{T}\!_\mathrm{f} \right\Vert\begin{cases}
< T_\mathrm{s} & \quad \text{for sticking,}\\
= T_\mathrm{s}  &  \quad \text{for sliding,}
\end{cases}
\end{equation}
and is computed using a predictor-corrector algorithm~\cite{Gouravaraju2020a}. A Neo-Hookean material model is employed to model the spatula response~\cite{Bonet2008}. For further details on the application of the adhesive friction model, we refer to Gouravaraju et al.~\cite{Gouravaraju2020a}.

\begin{figure}[h!]
	\begin{center}
		\includegraphics[scale=0.3]{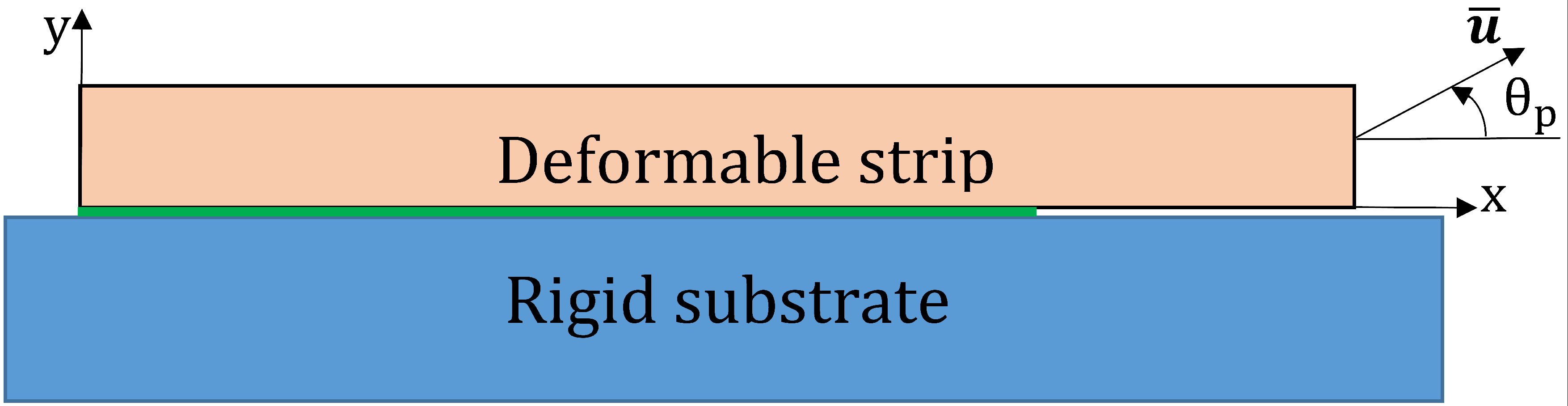}
		\caption{Peeling of a deformable strip from a rigid substrate. The strip is adhering on $75\%$ of the surface.}
		\label{fig:strip}
	\end{center}
\end{figure}

The spatula is modeled as a thin two-dimensional strip as shown in Fig.~\ref{fig:strip}. A displacement $\bar{\boldsymbol{u}}$ is applied to the spatula shaft at an angle called the peeling angle $\theta_\mathrm{p}$. Nonlinear finite element analysis is employed to solve the resulting mechanical boundary value problem given by the nonlinear equation
\begin{equation}
\mathbf{f}(\mathbf{u}):= \mathbf{f}_\mathrm{int} + \mathbf{f}_\mathrm{c} = \mathbf{0}\,,
\label{eq:EQM}
\end{equation}
where $\mathbf{f}_\mathrm{int}$ and $\mathbf{f}_\mathrm{c}$ are the global internal and contact force vectors. The spatula is divided into $240\times12$ finite elements along x and y directions, respectively. To accurately capture the nonlinear contact tractions (see Eqs.~(\ref{eq:sl_thresh}) and (\ref{eq:Adhesive_Traction_Ref})), a local enrichment strategy proposed by Sauer~\cite{RogerEnriched2011} is employed. In this strategy, the contact surface is discretized using fourth-order Lagrange polynomials while the bulk is discretized using the standard linear Lagrange polynomials. Plane strain conditions are assumed.

\begin{figure}[h!]
	\begin{center}
		\includegraphics[scale=0.32]{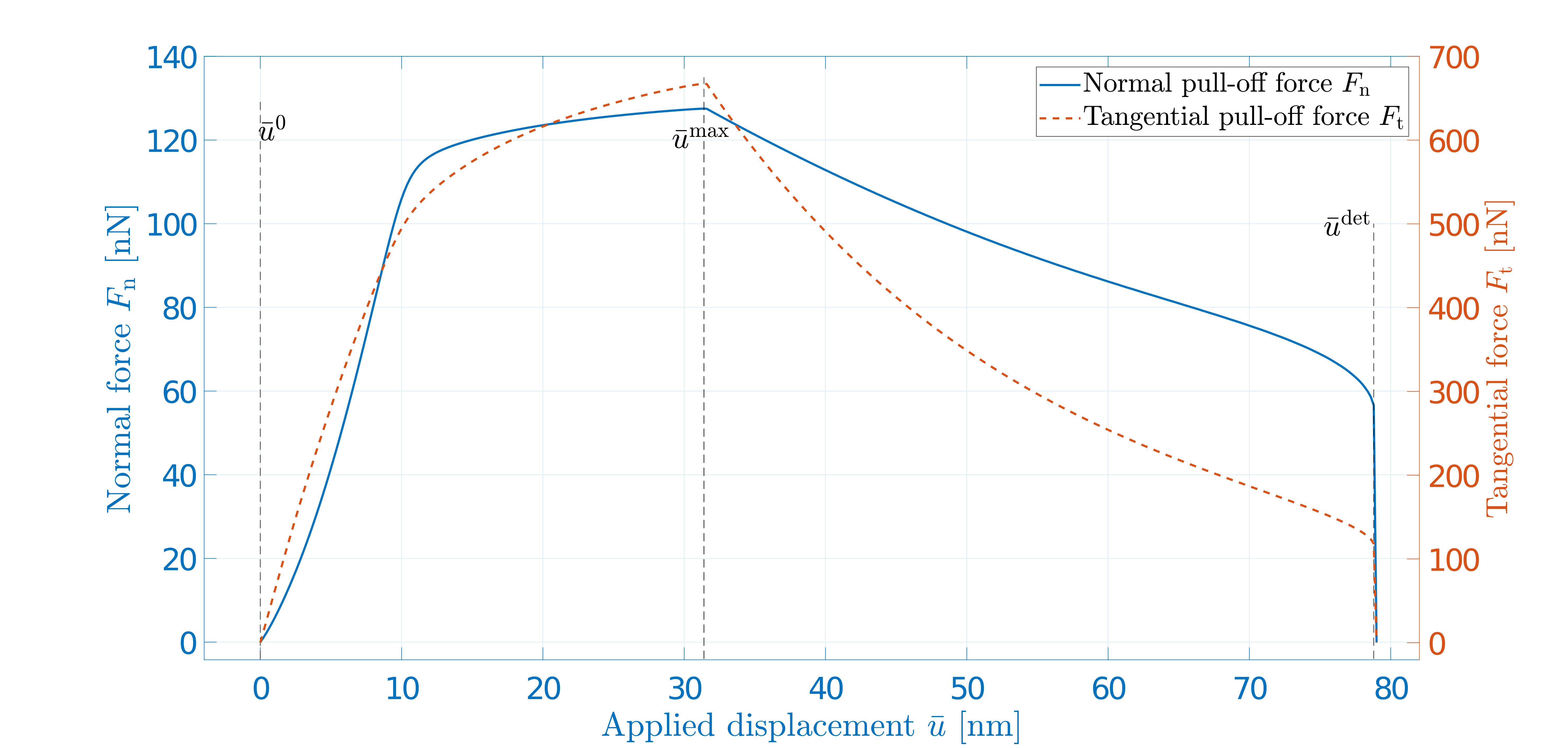}
		\caption{Evolution of normal ($F_\mathrm{n}$) and tangential ($F_\mathrm{t}$) pull-off forces with the applied displacement $\bar{u}$ for peeling angle $\theta_\mathrm{p} = 45^\circ$.}
			\label{fig:Peel_phases}
	\end{center}
\end{figure}

Although the detailed results of the FE simulation can be found in  Gouravaraju et al.~\cite{Gouravaraju2020a,Gouravaraju2020b}, for the sake of completeness, we briefly discuss the peeling process through a representative force-displacement plot. The entire peeling of the spatula can be divided into two phases based on the evolution of the normal and tangential pull-off forces shown in Fig.~\ref{fig:Peel_phases}. In the first phase (from displacement $\bar{u}^0$ to $\bar{u}^\mathrm{max}$), the spatula continuously undergoes stretching due to the fact that it is in a state of partial sliding/sticking near the peeling front. Thus, it accumulates strain energy. At $\bar{u}^\mathrm{max}$ the spatula is stretched to the maximum as the pull-off forces reach a maximum value. During the second phase (from $\bar{u}^\mathrm{max}$ to $\bar{u}^\mathrm{det}$) the spatula fully slides on the substrate. As a result, the spatula relaxes and releases the accumulated energy until it detaches from the substrate spontaneously at $\bar{u}^\mathrm{det}$. Similar peeling curves are obtained for other peeling angles.

In this study, the focus is on three aspects of the peeling process, viz.~the maximum normal pull-off force $F_\mathrm{n}^\mathrm{max}$, the maximum tangential pull-off force $F_\mathrm{t}^\mathrm{max}$, and the resultant force angle $\alpha = \arctan(F_\mathrm{n}/F_\mathrm{t})$ at detachment. It has been shown that depending on the peeling angle $\theta_\mathrm{p}$, the maximum pull-off forces $F_\mathrm{n}^\mathrm{max}$ and $F_\mathrm{t}^\mathrm{max}$, the corresponding displacement $\bar{u}^\mathrm{max}$ and the detachment displacement $\bar{u}^\mathrm{det}$ vary considerably~\cite{Gouravaraju2020a}. On the other hand, it has been observed  ~\cite{Gouravaraju2020a,Gouravaraju2020b} that the resultant force angle at detachment $\alpha^\mathrm{det}$ remains the same irrespective of the peeling angle (see Table~\ref{tab:FEdata} in Appendix~\ref{app:A}).

\emph{Remark 1}: Note that the geometrical and material parameters are fixed for the case of gecko spatula peeling, see ~\cite{Gouravaraju2020a,Gouravaraju2020b}. Hence, the effect of variation of these parameters is not considered. However, the effect of the geometrical or physical parameters can be incorporated by generating additional FE data and retraining the proposed network with the additional parameters added as input.

\section{Bayesian regularization-backpropagation neural network (BR-BPNN)}
In this section, a backpropagation neural network (BPNN) along with the Bayesian regularization learning algorithm is described.  The background theory on BPNN along with the Bayesian regularization is given in Appendix~\ref{app:brbpnn_theory_initial}. A more detailed discussion can be found in Demuth et al.~\cite{Hagan2014}. BR-BPNN is utilized to achieve better generalization and minimal over-fitting for the trained networks~\cite{MacKay1992,Burden2008}.

Consider a neural network with training dataset $D$ having $n_t$ input  and target  vector pairs in the network model, i.e
\begin{equation}
D = \Big\{\left(\mathbf{u}_1,\mathbf{t}_{\mathrm{o}1}\right), \left(\mathbf{u}_2,\mathbf{t}_{\mathrm{o}2}\right), \ldots, \big(\mathbf{u}_{n_t},\mathbf{t}_{\mathrm{o}n_t}\big)\Big\}\,.
\label{eq:define_D}
\end{equation}

For each input ($\mathbf{u}$) to the network, the difference between target output ($\mathbf{t}_\mathrm{o}$) and predicted output ($\mathbf{a}_\mathrm{o}$) is computed as error $\mathbf{e}$. In order to evaluate the performance of the network, i.e. how well the neural network is fitting the test data, a quantitative measure is needed. This measure is called performance index of the network and is used to optimize the network parameters. The standard performance index $F(\bar{\mathbf{w}})$ is governed by the sum of the squared errors (SSE)
\begin{equation}
F(\bar{\mathbf{w}}) = E_D = \sum_{i=1}^{n_t} \left(\mathbf{e}_i\right)^2= \sum_{i=1}^{n_t} \left(\mathbf{t}_{\mathrm{o}i} - \mathbf{a}_{\mathrm{o}i}\right)^\mathrm{T}\,\left(\mathbf{t}_{\mathrm{o}i} - \mathbf{a}_{\mathrm{o}i}\right)\,,
\label{eq:ED}
\end{equation}
where $\bar{\mathbf{w}}$ denotes the vector of size $K$ containing all the weights and biases of the network.

In order to generalize the neural network, the performance index of Eq.~(\ref{eq:ED}) is modified using a regularization method. A penalty term $(\mu/\nu) E_\mathrm{w}$ is added to the performance index $F(\bar{\mathbf{w}})$~\cite{Tikhonov1963},
\begin{equation}
F\big(\bar{\mathbf{w}}\big) = \mu \bar{\mathbf{w}}\!^\mathrm{T}_{}\, \bar{\mathbf{w}} + \nu E_D \,=\, \mu E_\mathrm{w} + \nu E_D \,,
\label{eq:obj_func}
\end{equation}
where $\mu$ and $\nu$ are the regularization parameters and $E_\mathrm{w}$ represents the sum of the squared network weights (SSW).

Finding the optimum values for $\mu$ and $\nu$ is a challenging task, as their comparative values set up the basis for the training error. If $\mu \,\ll \, \nu$, smaller errors are generated, while if $\mu \,\gg \, \nu$, there should be reduced weight size at the cost of network errors~\cite{Kayri2016}. For the purpose of finding the optimum regularization parameters, a Bayesian regularization method is employed.

Considering the network weights $\bar{\mathbf{w}}$ as random variables, the aim is to choose the weights that maximize the posterior probability distribution of the weights $P\big(\bar{\mathbf{w}}|D, \mu, \nu, M\!_N\big)$ given a certain data $D$. According to Bayes' rule~\cite{MacKay1992}, the posterior distribution of the weights depends on the likelihood function $P\big(D|\bar{\mathbf{w}},\nu ,M\!_N\big)$, the prior density $P\big(\bar{\mathbf{w}}|\mu ,M\!_N\big)$, and the normalization factor $P\big(D|\mu ,\nu ,M\!_N\big)$ for a particular neural network model $M_{N}$ and can be evaluated from
\begin{equation}
P\big(\bar{\mathbf{w}}|D, \mu, \nu, M\!_N\big) = \dfrac{P\big(D|\bar{\mathbf{w}}, \nu, M\!_N\big) \, P\big(\bar{\mathbf{w}}|\mu, M\!_N\big)}{P\big(D|\mu, \nu, M\!_N\big)}\,.
\label{eq:Bayes1}
\end{equation}

Considering that the noise in the training set has a Gaussian distribution, the likelihood function is given by
\begin{equation}
P\big(D|\bar{\mathbf{w}},\nu,M\!_N\big) = \frac{\exp\!\big(-\nu E_D\big)}{Z_D\big(\nu\big)}\,,
\end{equation}
where $Z_D = \big(\pi/\nu\big)^{Q/2}$ and $Q = n_t \times N^{n_l}$.

Similarly, assuming a Gaussian distribution for the network weights, the prior probability density $P\big(\bar{\mathbf{w}}|\mu ,M\!_N\big)$ is given as
\begin{equation}
P\big(\bar{\mathbf{w}}|\mu ,M\!_N\big) = \frac{\exp\!\big(-\mu E_\mathrm{w}\big)}{Z_\mathrm{w}\big(\mu\big)}\,,
\end{equation}
where $Z_\mathrm{w} = \big(\pi/\alpha\big)^{K/2}$.

The posterior probability with the network weights $\bar{\mathbf{w}}$ can then be expressed as \cite{Kayri2016}
\begin{equation}
P\big(\bar{\mathbf{w}}|D, \mu, \nu, M\!_N\big) = \dfrac{\exp\!\big(-\mu E_\mathrm{w} - \nu E_D\big)}{Z_F\big(\mu, \nu\big)} = \dfrac{\exp\!\big(-F(\bar{\mathbf{w}})\big)}{Z_F\big(\mu, \nu\big)}\,,
\label{eq:Bayes2}
\end{equation}
where $Z_F\big(\mu, \nu\big) = Z_D\big(\nu\big) Z_\mathrm{w}\big(\mu\big)$ is the normalization factor.

The complexity of the model $M_{N}$ is governed by regularization parameters $\mu$ and $\nu$, which need to be estimated from the data. Therefore, Bayes' rule is again applied to optimize them from
\begin{equation}
P\big(\mu, \nu|D, M\!_N \big) = \dfrac{P\big(D|\mu, \nu, M\!_N\big)\, P\big(\mu, \nu|M\!_N\big)}{P\big(D|M\!_N\big)}\,,
\label{eq:Bayes3}
\end{equation}
where $P\big(\mu, \nu|M\!_N \big)$ denotes the assumed uniform prior density for the parameters $\mu$ and $\nu$. From Eq.~(\ref{eq:Bayes3}), it is evident that maximizing the likelihood function $P\big(D|\mu, \nu, M\!_N\big)$ eventually maximizes the posterior probability $P\big(\mu,\nu|D, M\!_N \big)$. Moreover, it can be noted that the likelihood function in Eq.~(\ref{eq:Bayes3}) is the normalization factor of Eq.~(\ref{eq:Bayes1}). Therefore, solving for the likelihood function $P\big(D|\mu ,\nu ,M\!_N \big)$ and expanding the objective function in Eq.~(\ref{eq:obj_func}) around the minimal point $\bar{\mathbf{w}}^*$ via a Taylor series expansion, the optimum values of regularization parameters can be evaluated as follows~\cite{Foresee1997}
\begin{equation}
\mu^* = \dfrac{\gamma}{2 E_\mathrm{w}\big(\bar{\mathbf{w}}^*\big)} \quad \quad \mathrm{and} \quad \quad \nu^* = \dfrac{Q-\gamma}{2E_D\big(\bar{\mathbf{w}}^*\big)}\,,
\label{eq:reg_param}
\end{equation}
where $\gamma$ signifies the ``number" of effective parameters exhausted in minimizing the error function
\begin{equation}
\gamma = K-{\mu }^*\mathrm{tr}\big(\mathbf{H}^*\big)^{-1}, \text{for} \quad \quad 0 \leq \gamma \leq K\,, \label{eq:gamma}
\end{equation}
and $\mathbf{H}^*$ is the Hessian matrix of the objective function evaluated at $\bar{\mathbf{w}}^*$, which is calculated using the Gauss-Newton approximation as~\cite{Kayri2016}
\begin{equation}
\mathbf{H}^* \approx \mathbf{J}^\mathrm{T}\mathbf{J}\,,
\label{eq:Hessian}
\end{equation}
where $\mathbf{J}$ is the Jacobian matrix formed by the first derivatives of the network errors $\mathbf{e}$ with respect to network weights $\emph{w}_{ij}$. In (\ref{eq:gamma}), $\mathrm{tr}(\cdot)$ denotes the trace operator. The normalization factor $Z_F(\mu,\nu)$ can then be approximated as~\cite{Hagan2014}
\begin{equation}
Z_F\big(\mu, \nu\big) \approx \big(2\pi\big)^{K/2} \, \big(\mathrm{det}\big(\mathbf{H}^*\big)\big)^{-1/2}\, \exp\!\big(\!\!-F\big(\bar{\mathbf{w}}^*\big)\big)\,.
\end{equation}

At the end of the training, a few checks regarding the number of effective parameters are required for better performance of the network \cite{Kayri2016}. The problem of computing the Hessian matrix at the minimal point $\bar{\mathbf{w}}^*$ is implicitly solved in the Levenberg-Marquardt (LM) training algorithm while finding the minimum of $F(\bar{\mathbf{w}})$. In the LM algorithm, the network weights and biases at the $k^{th}$ iteration are adjusted according to ~\cite{MacKay1992,Foresee1997}
\begin{equation}
\bar{\mathbf{w}}^{k+1} = \bar{\mathbf{w}}^k - \big[\mathrm{\textbf{J}}^\mathrm{T} \mathrm{\textbf{J}} + \lambda \mathrm{\textbf{I}} \big]^{-1}\,\mathrm{\textbf{J}}^\mathrm{T} \mathbf{e}\,, \label{eq:weight_update}
\end{equation}
where $\lambda$ denotes the Levenberg's damping factor and $\mathrm{\textbf{J}}^\mathrm{T} \mathbf{e}$ is the error gradient, which needs to be close to zero at end of the training.

\section{Implementation of BR-BPNN}
In this work, the input vector $\mathbf{u}$ of the BR-BPNN models contains seventeen elements with peeling angle values $\theta_\mathrm{p}$ ranging from $10^\circ$ to $90^\circ$ at an interval of $5^\circ$. The corresponding output vectors are the maximum normal pull-off force $\boldsymbol{F}_\mathrm{n}^\mathrm{max}$, the maximum tangential pull-off force $\boldsymbol{F}_\mathrm{t}^\mathrm{max}$, the applied displacement at force maximum $\bar{\boldsymbol{u}}^\mathrm{max}$, the resultant force angle at detachment $\boldsymbol{\alpha}^\mathrm{det}$, and the applied displacement at detachment $\bar{\boldsymbol{u}}^\mathrm{det}$. In general, this input-output dataset is randomly divided into training, validation, and testing sub-datasets. The training dataset is used to train the neural network model (which in the current work is carried out using the Bayesian regularization-backpropagation method) and the trained model is further validated with the validation dataset.

As described previously, the neural network model is first trained on the training dataset and its performance is evaluated by making predictions using the testing dataset. However, this type of single-run model-validation method could potentially result in selection-bias, i.e. the accuracy of the model will be highly dependent on the particular choice of the training and testing datasets. In order to assess the effectiveness of a neural network model developed using limited data, as in this work, a cross-validation method called $k$-fold cross-validation method is employed in the training of neural networks. This helps the neural network to generalize to new or unseen data in a much better manner. In the $k$-fold cross-validation method  the complete dataset is divided into two complementary sub-datasets, i.e. training and testing.  In this method, for a given neural network model, the dataset is first randomized and then partitioned (split) in to  $k$ almost equal sized sub-datasets called folds. Then, the $k-1$ folds are used to train the neural network. The one remaining  fold (i.e., $k^{\mathrm{th}}$ fold) is used for testing the performance of the neural network model. This process is repeated $k$ times such that the network is trained and tested on the entire dataset as illustrated in Table~\ref{tab:kfold} which shows the dataset split five times (Split 1 to Split 5)  into five folds (Fold 1 to Fold 5). The yellow cells in Table~\ref{tab:kfold} represent testing dataset while the blue cells correspond to training dataset. The performance of the neural network is then reported in terms of the average accuracy obtained from this $k$-fold cross-validation.
\begin{center}
\fontsize{9}{11}\selectfont{
\begin{table}
\centering
	\caption{$K$-fold cross-validation with dataset split into five folds. The yellow cells represent testing dataset while the blue cells correspond to training dataset. See Table~\ref{tab:kfold_details} for explicit details for the folds used in the present work.}\label{tab:kfold}
	\begin{tabular}{|l|c|c|c|c|c|}
		\hline
		\textbf{Split 1} & \cellcolor{yellow} Fold 1  & \cellcolor{blue!35} Fold 2 & \cellcolor{blue!35} Fold 3 & \cellcolor{blue!35} Fold 4 & \cellcolor{blue!35} Fold 5\\
		\hline
		\textbf{Split 2} & \cellcolor{blue!35} Fold 1 &\cellcolor{yellow} Fold 2 & \cellcolor{blue!35} Fold 3 & \cellcolor{blue!35} Fold 4 & \cellcolor{blue!35} Fold 5\\
		\hline
		\textbf{Split 3} & \cellcolor{blue!35} Fold 1 &\cellcolor{blue!35} Fold 2 & \cellcolor{yellow} Fold 3 & \cellcolor{blue!35} Fold 4 & \cellcolor{blue!35} Fold 5\\
		\hline
		\textbf{Split 4} & \cellcolor{blue!35} Fold 1 & \cellcolor{blue!35} Fold 2 & \cellcolor{blue!35} Fold 3 & \cellcolor{yellow} Fold 4 & \cellcolor{blue!35} Fold 5\\
		\hline
		\textbf{Split 5} & \cellcolor{blue!35} Fold 1 &\cellcolor{blue!35} Fold 2 & \cellcolor{blue!35} Fold 3 & \cellcolor{blue!35} Fold 4 & \cellcolor{yellow} Fold 5\\
		\hline
	\end{tabular}
\end{table}
}
\end{center}
\begin{center}
\fontsize{9}{11}\selectfont{
\begin{table}
\centering
	\caption{Details of  training dataset  and testing dataset used in the $k$-fold cross-validation. The indices refer to the case number (first column) in  Table~\ref{tab:FEdata}.}\label{tab:kfold_details}
	\begin{tabular}{|l|p{1.6in}|p{0.7in}|p{0.9in}|p{0.7in}|p{0.7in}|}
		\hline
         Split Number & Training  dataset indices & Number of training data ($N_{\mathrm{train}}$)  & Testing dataset indices &  Number of testing data ($N_{\mathrm{test}}$) & Fold for Testing dataset\\
         \hline
		\textbf{Split 1} &  1, 2, 5, 6, 8, 10, 11, & 13 & 3, 4, 7, 9 & 4 & Fold 1 \\
                         &  12, 13, 14, 15, 16, 17 & & & & \\
		\hline
		\textbf{Split 2} & 1, 2, 3, 4, 5, 6, 7, 8, 9, & 14 & 12, 13, 17 & 3 &Fold 2 \\
                         &   10, 11, 11, 14, 15, 16 & & & & \\
		\hline
		\textbf{Split 3} &  2, 3, 4, 5, 7, 8, 9, 12, & 13 & 1, 6, 10, 11 &  4 & Fold 3 \\
                         &  13, 14, 15, 16, 17 & & & & \\
		\hline
		\textbf{Split 4} & 1, 3, 4, 5, 6, 7, 9, 10, &  14 & 2, 8, 14 & 3 & Fold 4 \\
                         &  11, 12, 13, 15, 16, 17 & & & & \\
		\hline
		\textbf{Split 5} & 1, 2, 3, 4, 4, 6, 7, 8, 9 &  14 & 5, 15, 16 & 3 & Fold 5 \\
                         & 10, 11, 12, 13, 14, 17 & & & & \\
		\hline
	\end{tabular}
	
\end{table}
}
\end{center}

Table~\ref{tab:kfold_details} give the details of  testing dataset  and testing dataset used in the $k$-fold cross-validation, see Appendix~\ref{app:A} for the FE results. The indices in the table refer to the case number in Table~\ref{tab:FEdata} (first column). For each split, the training dataset is used to train the neural network model using Bayesian regularization method and the trained model is further validated with the validation dataset using the fold mentioned in the last column of Table~\ref{tab:kfold_details}.  The validation dataset, in other back-propagation training algorithms, is used to optimize the hyperparameters for effective training. The hyperparameters, like the number of neurons in the hidden layer and the learning parameters such as $\gamma$ and $\lambda$, are defined as the variables required for training the neural network. However, for BR-based learning networks, the hyperparameters in the form of the regularization parameters ($\mu, \nu $) are implicitly optimized using Eq.~(\ref{eq:obj_func}). Therefore, the validation set is not essentially required in this case for optimizing the network hyperparameters. Finally, the testing dataset is utilized to predict the targeted output $\mathbf{t}_\mathrm{o}$ and analyze the model performance, accordingly. Appendix~\ref{app:B} presents a simple algorithmic overview of the BR-BPNN model developed in the present work.

Next, two BR-BPNN models are formed with different output datasets; the first model has three output vectors and the second model has two output vectors as shown in Tables~\ref{tab:out_data_1} and ~\ref{tab:out_data_2}. The three output vectors for BR-BPNN-I are the applied displacement at force maximum $\bar{\boldsymbol{u}}^\mathrm{max}$, the maximum normal pull-off force $\boldsymbol{F}_\mathrm{n}^\mathrm{max}$, and the maximum tangential pull-off force $\boldsymbol{F}_\mathrm{t}^\mathrm{max}$.  For BR-BPNN-II, the output vectors are the applied displacement at detachment $\bar{\boldsymbol{u}}^\mathrm{det}$ and the resultant force angle at detachment $\boldsymbol{\alpha}^\mathrm{det}$, respectively. Each output vector consists of $3 N_{\mathrm{test}}$ and $2 N_{\mathrm{test}}$ elements for models BPNN-I and BPNN-II respectively.

However, only $N_{\mathrm{train}}$ elements corresponding to the input training dataset (see Table~\ref{tab:kfold_details}) are selected for training the BPNN models. Then, the input and output vectors are normalized by the corresponding maximum values. The performance of the BR-BPNN models are estimated by comparing the mean square error (MSE)  values with the number of neurons in the hidden layer and determining the optimal number.  The MSE is computed from the network error $E_D$ in Eq.~(\ref{eq:ED}) as
\begin{equation}
\mathrm{MSE}  = \frac{1}{n_t} E_D. \label{eq:MSE}
\end{equation}
\begin{table}
	\caption{Output dataset for model BR-BPNN-I (see Appendix~\ref{app:A} for the FE results).}\label{tab:out_data_1}
	\begin{tabular}{ p{3in}  p{3in} }
		\hline \\
         Applied displacement at force maximum & $\bar{\boldsymbol{u}}^\mathrm{max} := \left[\bar{u}_1^\mathrm{max}, \bar{u}_2^\mathrm{max}, \ldots \ldots, \bar{u}_{16}^\mathrm{max}, \bar{u}_{17}^\mathrm{max}\right]^\mathrm{T}$  \\ \\
		Maximum normal pull-off force &  $\boldsymbol{F}_\mathrm{n}^\mathrm{max} := \left[F_{n_1}^\mathrm{max}, F_{n_2}^\mathrm{max} \ldots \ldots, F_{n_{16}}^\mathrm{max}, F_{n_{17}}^\mathrm{max}\right]^\mathrm{T}$    \\ \\
		Maximum tangential pull-off force & $\boldsymbol{F}_\mathrm{t}^\mathrm{max} :=\left[F_{t_1}^\mathrm{max}, F_{t_2}^\mathrm{max} \ldots \ldots, F_{t_{16}}^\mathrm{max}, F_{t_{17}}^\mathrm{max}\right]^\mathrm{T}$   \\ \\
		\hline
	\end{tabular}
\end{table}

\begin{table}
	\caption{Output dataset for model BR-BPNN-II (see Appendix~\ref{app:A} for the FE results).}\label{tab:out_data_2}
	\begin{tabular}{ p{3in} p{3in}}
		\hline \\
         Applied displacement at detachment & $\bar{\boldsymbol{u}}^\mathrm{det} :=\left[\bar{u}^\mathrm{det}_1, \bar{u}^\mathrm{det}_2, \ldots \ldots, \bar{u}^\mathrm{det}_{16}, \bar{u}^\mathrm{det}_{17}\right]^\mathrm{T}$    \\ \\
		Resultant force angle at detachment &  $\boldsymbol{\alpha}^\mathrm{det} :=\left[\alpha^\mathrm{det}_1, \alpha^\mathrm{det}_2 \ldots \ldots, \alpha^\mathrm{det}_{16}, \alpha^\mathrm{det}_{17}\right]^\mathrm{T}$   \\ \\
		\hline
	\end{tabular}
\end{table}

\emph{Remark 2}: Even though the geometric and material parameters for gecko spatula peeling are cosidered fixed, see Remark 1 at the end of Section 2, the proposed model can be extended to predict the influence of these parameters as follows: First an additional FE dataset needs to be generated for each parameter. Then, the input vector of the proposed model needs to be extended to include the additional input parameters. The network can then be retrained to obtain the optimum number of neurons in the hidden layer and the model parameters. The algorithm mentioned in appendix~\ref{app:B} will then, in principle, work in a similar manner.

\emph{Remark 3}: It is worth noting that in this work a neural network-based prediction of adhesion phenomena is proposed rather than applying a curve-fitting-based interpolation technique. At first it may appear that the proposed BR-BPNN models merely interpolate the missing data. However, this is not so due to the following reasons:
\begin{itemize}
  \item It can be observed from Table~\ref{tab:FEdata} that all the output vectors consist of high dimensional data and the change from preceding value to the next one is highly nonlinear. It is well known that for the case of highly nonlinear data, neural networks can provide more flexibility in mapping the input-output relation with accurate tolerances. Moreover, in case more precise results are desired using curve fitting, the selection of high dimensional polynomials increases the computational complexity and eventually the computational time. This is a major drawback of curve fitting.
  \item Furthermore, even if curve fitting can be used for interpolation, it is pertinent to mention that the generalization capability of the curve fitting technique can not be as accurate as the proposed BR-BPNN when the dimensionality of the data increases. The proposed BR-BPNN models perform well because the procedure to map the input-output dataset is inherently interpreted by the systematic selection of activation function, hyperparameters, neurons, and hidden layer(s). However, in case of curve-fitting, this process is an iterative one left as a user input for the selection of a polynomial function.
\end{itemize}

\emph{Remark 4}: It is also worth to mention that the computational time taken for each full finite element run shown in Table~\ref{tab:FEdata}, depending on the peeling angle value, takes between $15$ minutes to $7$ hours on a multicore machine with parallel computing. On the other hand the training and the testing of the dataset in Table~\ref{tab:kfold_details} takes less than a minute for each split on the same machine without any parallel computing option enabled.

\section{Results and discussion}\label{results_discussions}
This section presents the Bayesian regularization-based backpropagation neural network predictions of the maximum normal pull-off force $F_\mathrm{n}^\mathrm{max}$, the maximum tangential pull-off force $F_\mathrm{t}^\mathrm{max}$, and the resultant force angle at detachment $\alpha^\mathrm{det}$ along with the corresponding displacements $\bar{u}^\mathrm{max}$ and $\bar{u}^\mathrm{det}$. Predictions of the networks are then compared with the FE results of Gouravaraju et al.~\cite{Gouravaraju2020a,Gouravaraju2020b} that have not been yet used for training.
\begin{figure}[h!]
	\centering
	\includegraphics[scale=.3]{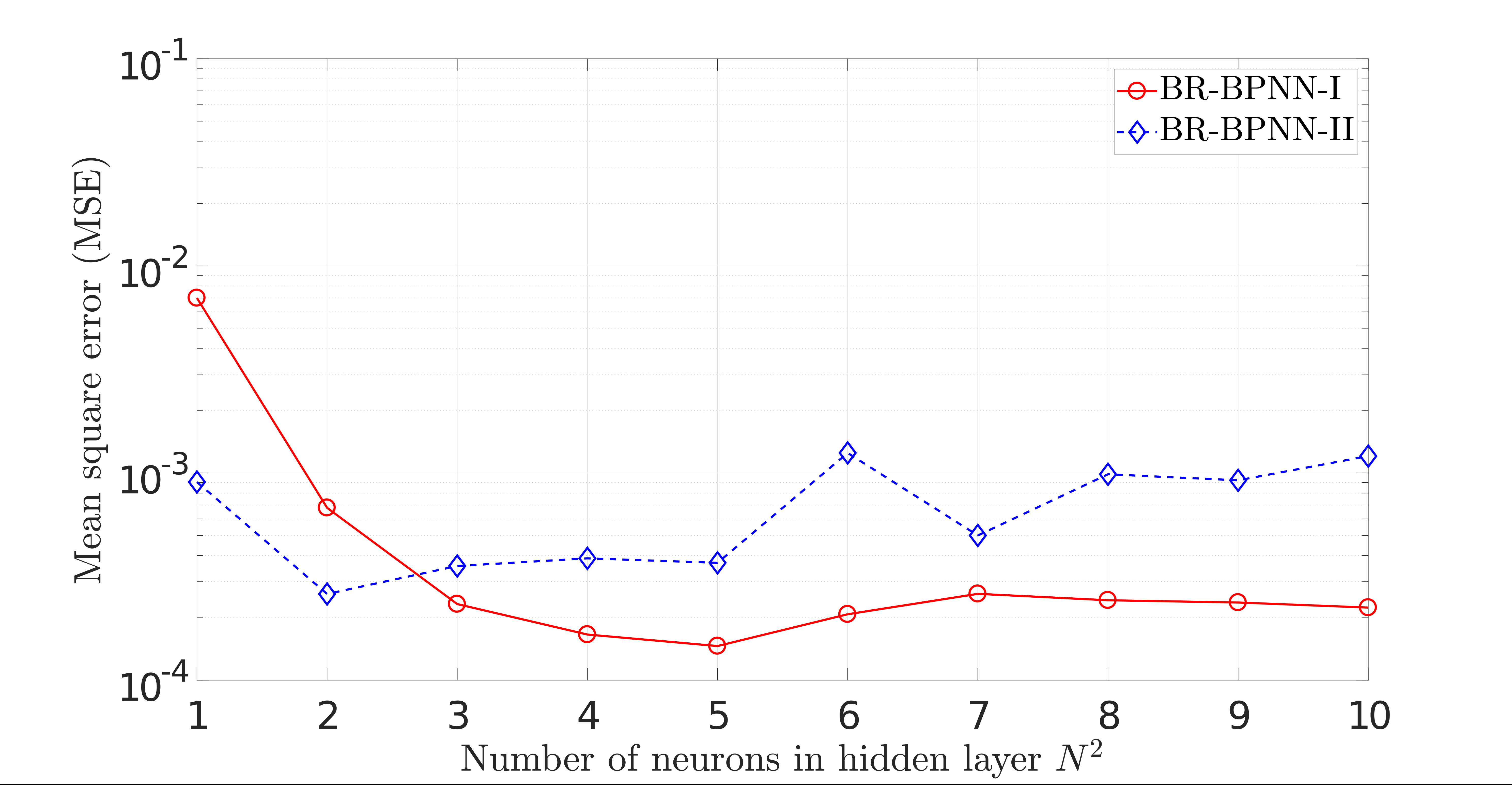}
	\caption{Average mean square error from 5-fold cross-validation with the number of neurons in the hidden layer for different BR-BPNN models.}
	\label{fig:MSE}
\end{figure}

To define the optimal structure of each network model, the mean square error (MSE) of Eq.~(\ref{eq:MSE}) is investigated along with the number of neurons ($1$ to $10$) in the hidden layer. For the two BR-BPNN models (BR-BPNN-I and BR-BPNN-II ), training is performed with $1$ to $10$ hidden neurons. The MSE values for both the models with only one hidden neuron are found to be comparatively high i.e. $7 \times 10^{-3}$ and $9.045 \times 10^{-4}$, being incapable to form an efficient network. However, as the number of hidden neurons increases to two, a major drop in the MSE values ($6.793 \times 10^{-4}$, and $2.6060 \times 10^{-4}$) is recorded. Each model is trained $15$ times independently for different number of neurons to mitigate the unfavorable effects by choosing random initial weights. Each network model is trained for a maximum of $2000$ epochs. An epoch is completed when the entire training dataset is passed forward and backward through the network thus updating the weights once. For BPNN-I, the mean square error attains a broad minimum and continuous to decrease between $1$ and $5$ hidden neurons as shown in Fig.~\ref{fig:MSE}. For $N^2$ greater than $5$, the MSE value again starts to rise due to overfitting of the network models. Therefore, for BPNN- I the number of neurons in the hidden layer is selected as $5$. The number of neurons in the input and output layers are taken as $1$ and $3$ as there is one input vector and three output vectors for the BPNN-I model. Following a similar trend, the optimal number of hidden neurons for model BPNN-II is found to be $2$, forming the network structure $1$-$2$-$2$.

Either of the following criteria are selected to terminate or complete the training process: maximum number of epochs reached, minimum value of performance gradient reached, minimum constant value of effective parameters ($\gamma$) reached, maximum value of Levenberg's damping factor ($\lambda$) attained, or MSE reaching the performance limits. The training results for models BR-BPNN-I and BR-BPNN-II are shown in Tables~\ref{tab:TP_BRBPNN1} and ~\ref{tab:TP_BRBPNN2} respectively. The other network training parameters like the sum of square errors (SSE) (Eq.~(\ref{eq:ED})), sum of square weights (SSW) ($ E_\mathrm{w}$ in Eq.~(\ref{eq:obj_func})), Levenberg's damping factor, and error gradient (Eq.~(\ref{eq:weight_update})) values are also shown in Tables~\ref{tab:TP_BRBPNN1} and ~\ref{tab:TP_BRBPNN2}.
\begin{table}
\centering
	\caption{Training parameters for the best configuration (1-5-3) for BR-BPNN-I from 5-fold cross validation.}\label{tab:TP_BRBPNN1}
\begin{tabular}{|p{0.5in}|p{0.5in}|p{0.84in}|p{0.84in}|p{0.4in}|p{0.7in}|p{0.84in}|p{0.84in}|} \hline
        & Epochs & MSE & SSE($E_D$) & SSW ($E_W$) & No. of effective parameters ($\gamma $) & LM Parameter ($\lambda $) & Gradient
(${\boldsymbol{\mathrm{J}}}^{\mathrm{T}}\boldsymbol{\mathrm{e}}$) \\ \hline
Split 1  & 384 & 9.79 $\times 10^{-4}$ & 3.26 $\times 10^{-4}$ & 63.98 & 22.18 & 1.0  & 9.88 $\times 10^{-8}$ \\ \hline
Split 2 & 300 & 7.39 $\times 10^{-5}$ & 2.47 $\times 10^{-5}$ & 80.97 & 22.38 & 1.0 $\times 10^{10}$ & 1.07 $\times 10^{-7}$ \\ \hline
Split 3 & 145 & 2.36 $\times 10^{-6}$ & 7.81 $\times 10^{-7}$ & 56.61 & 23.26 & 1.0 $\times 10^{10}$ & 1.31 $\times 10^{-7}$ \\ \hline
Split 4 & 92 & 1.86 $ \times 10^{-6}$ & 6.21 $\times 10^{-7}$ & 110.29 & 24.90 & 1.0 $\times 10^{10}$ & 1.34 $\times 10^{-7}$ \\ \hline
Split 5 & 106 & 0.0011 & 3.62 $\times 10^{-4}$ & 46.91 & 23.15 & 1.0 & 9.88 $\times 10^{-8}$ \\ \hline
\end{tabular}
\end{table}
\begin{table}
\centering
	\caption{Training parameters for best configuration (1-2-2) for BR-BPNN-II from 5-fold cross validation.}\label{tab:TP_BRBPNN2}
\begin{tabular}{|p{0.5in}|p{0.5in}|p{0.84in}|p{0.84in}|p{0.4in}|p{0.7in}|p{0.84in}|p{0.84in}|} \hline
 & Epochs & MSE & SSE($E_D$) & SSW ($E_W$) & No. of effective parameters ($\gamma $) & LM Parameter ($\lambda $) & Gradient
(${\boldsymbol{\mathrm{J}}}^{\mathrm{T}}\boldsymbol{\mathrm{e}}$) \\ \hline
Split 1  & 64 & 1.18 $\times 10^{-5}$ & 3.93 $\times 10^{-6}$ & 53.71 & 8.57 & 1.0 $\times 10^{10}$ & 4.23 $\times 10^{-7}$ \\ \hline
Split 2 & 82 & 1.27 $\times 10^{-5}$ & 4.24 $\times 10^{-6}$ & 60.20 & 8.26 & 1.0 $\times 10^{10}$& 5.76 $\times 10^{-7}$ \\ \hline
Split 3 & 43 & 8.38 $\times 10^{-6}$ & 2.79 $\times 10^{-6}$ & 37.29 & 8.31 & 1.0 $\times 10^{10}$ & 7.75 $\times 10^{-7}$ \\ \hline
Split 4 & 68 & 5.84 $\times 10^{-6}$ & 1.95 $\times 10^{-6}$ & 43.26 & 8.51 & 1.0 $\times 10^{10}$ & 8.93 $\times 10^{-7}$ \\ \hline
Split 5 & 115 & 8.94 $\times 10^{-6}$ & 2.98 $\times 10^{-6}$ & 62.13 & 8.43 & 1.0 $\times 10^{10}$ & 7.59 $\times 10^{-7}$ \\ \hline
\end{tabular}
\end{table}

After training the models with input-output datasets with $N_{\mathrm{train}}$ datapoints (see Table~\ref{tab:kfold_details}), the testing dataset with $N_{\mathrm{test}}$ datapoints (see Table~\ref{tab:kfold_details}) is utilized to predict the corresponding desired output values. The relative error (RE) is used to measure the accuracy of the network predictions. The RE is calculated as the deviation of the predicted result from the desired target result, i.e.
\begin{equation}
\mathrm{RE} = \frac{t_i - a_i}{t_i} \,,
\end{equation}
where $t_i$ and $a_i$ denote the desired target result and the network prediction for a particular peeling angle of the testing data set, respectively.

\subsection{Case I: Maximum normal and tangential pull-off forces}\label{results_discussion_case_I}
\begin{figure}[h!]
	\centering
	\includegraphics[scale=.3]{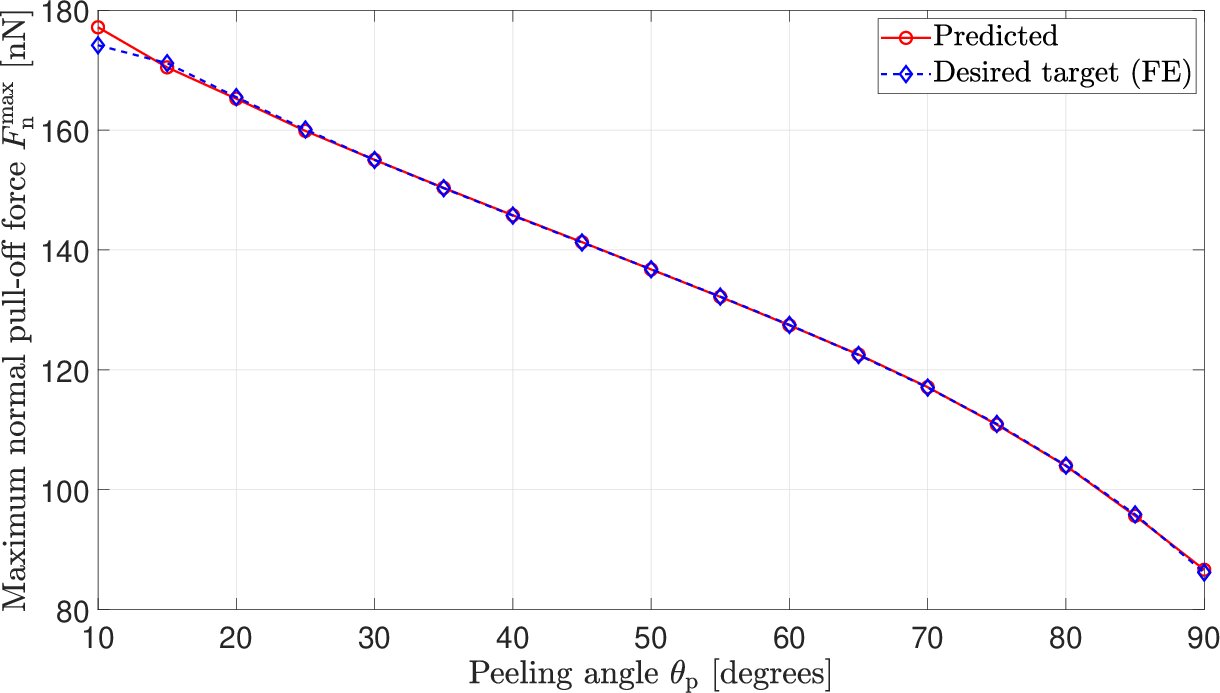}
	\caption{Plot of predicted and desired (FE) value of maximum normal pull-off force $F_\mathrm{n}^\mathrm{max}$ with the peeling angle $\theta_\mathrm{p}$ across all the splits for model BR-BPNN-I. }
	\label{fig:BPNN1_peeling_angle_max_FN}
\end{figure}

\begin{figure}[h!]
	\centering
	\includegraphics[scale=.3]{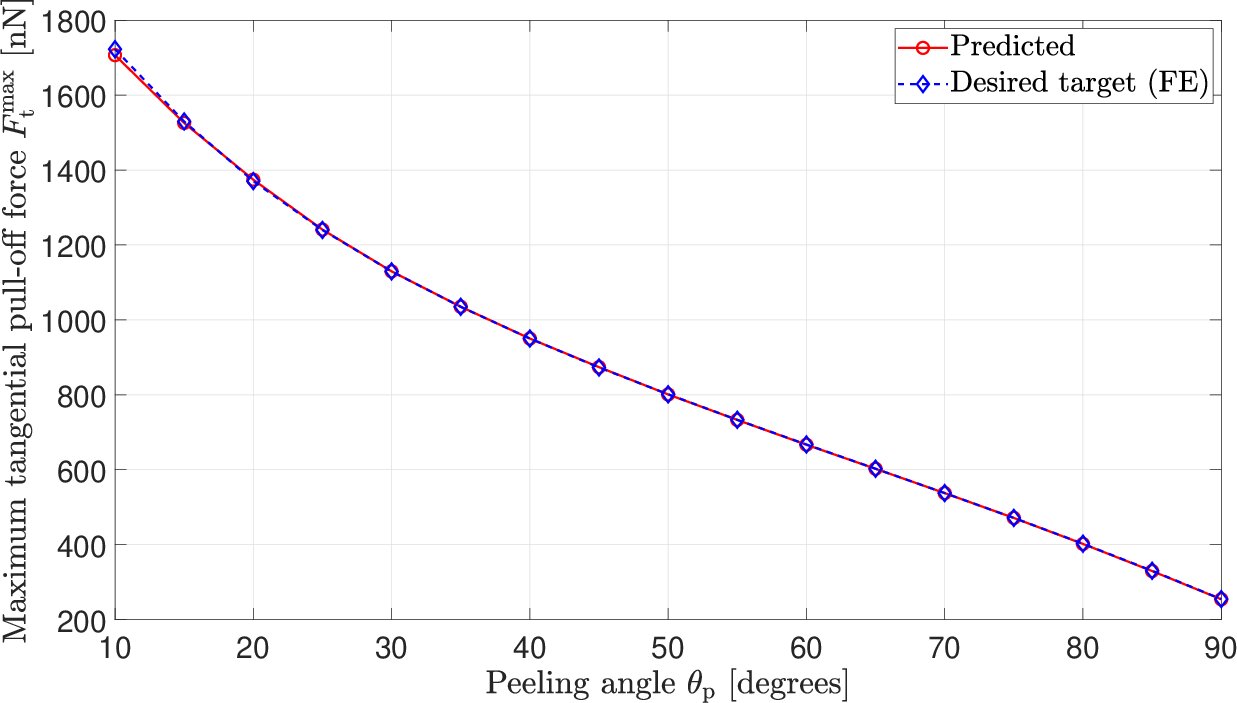}
	\caption{Plot of predicted and desired (FE) value of maximum tangential pull-off force $F_\mathrm{t}^\mathrm{max}$ with the peeling angle $\theta_\mathrm{p}$ across all the splits for model BR-BPNN-I.}
	\label{fig:BPNN1_peeling_angle_max_FT}
\end{figure}

\begin{figure}[h!]
	\centering
	\includegraphics[scale=.3]{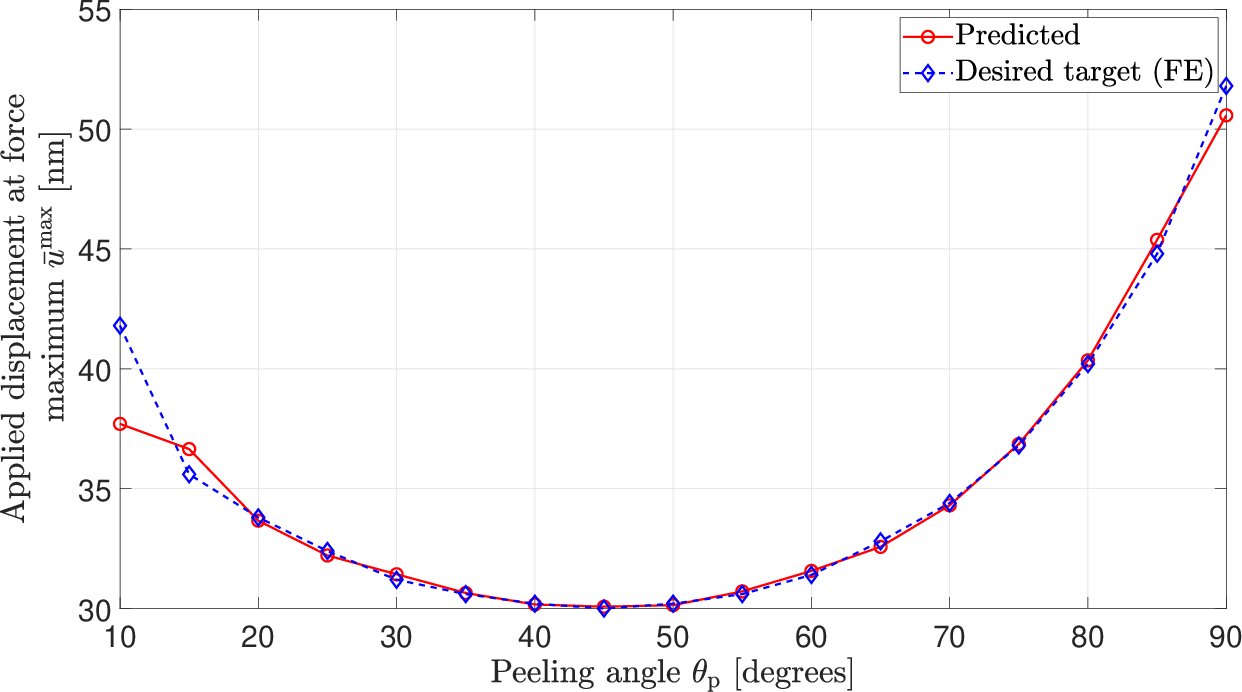}
	\caption{Plot of predicted  and desired (FE) value of applied displacement $\bar{u}^\mathrm{max}$ with the peeling angle $\theta_\mathrm{p}$ at maximum pull-off force  across all the splits for model BR-BPNN-I.}
	\label{fig:BPNN1_peeling_angle_max_U}
\end{figure}

Based on the training parameters from Table~\ref{tab:TP_BRBPNN1}, Figs.~\ref{fig:BPNN1_peeling_angle_max_FN},~\ref{fig:BPNN1_peeling_angle_max_FT}, and~\ref{fig:BPNN1_peeling_angle_max_U} present the predicted (BR-BPNN-I) results of the maximum normal pull-off force $F_\mathrm{n}^\mathrm{max}$, maximum tangential pull-off force $F_\mathrm{t}^\mathrm{max}$ and the corresponding applied displacement $\bar{u}^\mathrm{max}$. Since in the present work a 5-fold cross validation method is used the predicted and the desired results across all the splits are shown \footnote{ The correlation between the split and the predicted indices can be found in Table~\ref{tab:kfold_details}}. It can be seen from Figs.~\ref{fig:BPNN1_peeling_angle_max_FN} and~\ref{fig:BPNN1_peeling_angle_max_FT} that the predicted values of $F_\mathrm{n}^\mathrm{max}$ and $F_\mathrm{t}^\mathrm{max}$ for all angles except $\theta_\mathrm{p} = 10^\circ$ are very close to the desired target results (that are obtained by FE). However, for $\theta_\mathrm{p} = 10^\circ$, the predicted results show a slightly higher deviation compared to the other tested peeling angles.
It is observed from Table~\ref{tab:kfold_details} that the $10^\circ$ angle (index 1) is considered as part of the training dataset for the first, second, fourth, and fifth split and the testing dataset for the third split. Although the predicted results for $10^\circ$ angle are found to be more accurate for the third split, they, however, are computed for all the splits. Furthermore, as can be interpreted from Table~\ref{tab:TP_BRBPNN1}, the MSE values significantly contribute to those splits which have $10^\circ$  angle (index 1) in the training dataset. The MSE value in the third split can not compensate the adverse effects of the MSE values in rest of the splits. Therefore, this cumulative effect of the MSE values has become instrumental in creating  a disparity between the predicted results from the testing dataset and the desired results from the FE model. In the case with  $20^\circ$   angle, the disparity is substantially reduced by compensating the adverse effects of the training splits (first, second, third, and fifth) through the testing split (i.e., the fourth split) which has the smallest  MSE value.
The predictions are a little different for $\bar{u}^\mathrm{max}$ as shown in Fig.~\ref{fig:BPNN1_peeling_angle_max_U} where significant differences are found for $\theta_\mathrm{p} = 10^\circ , \, 20^\circ$, and $90^\circ$. This can also be observed from Table~\ref{tab:APE1}, which lists the relative error (RE) for the all the tested peeling  angles. From the table it can be seen that the maximum relative error for the case of displacement $\overline{u}^{\mathrm{max}}$ is $9.68$\% while the average relative error is around $1.22$\%. The average relative error for the case of maximum normal and tangential forces is found to be very small.

\begin{table}
\centering
	\caption{Relative error (RE) for the predictions of model BR-BPNN-I.}\label{tab:APE1}
\begin{tabular}{|p{1.5in}|p{1.1in}|p{1.1in}|p{1.1in}|} \hline
 & $F_\mathrm{n}^\mathrm{max}$ & $F_\mathrm{t}^\mathrm{max}$ & $\overline{u}^{\mathrm{max}}$ \\ \hline
Maximum RE (\%)  & $1.78$ & $0.91$ & $9.68$ \\ \hline
Minimum RE (\%) & $2.04\times {10}^{-4}$ & $0.0076$ & $0.06$ \\ \hline
Average RE (\%) & $0.24$ & $0.19$ & $1.22$ \\ \hline
\end{tabular}
\end{table}

\subsection{Case II: Resultant force angle at detachment}\label{results_discussion_case_II}
\begin{figure}[h!]
	\centering
	\includegraphics[scale=.3]{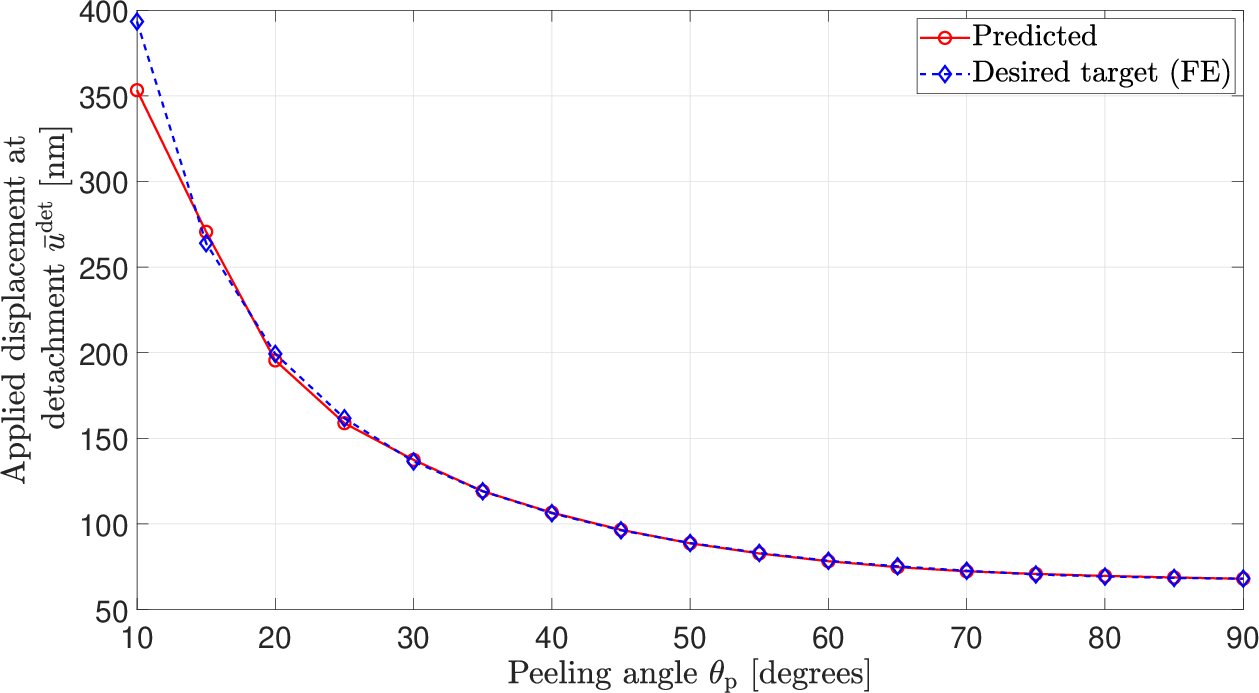}
	\caption{Plot of predicted and desired (FE) value of applied displacement at detachment $\bar{u}^\mathrm{det}$ with the peeling angle $\theta_\mathrm{p}$ across all the splits for model BR-BPNN-II. }
	\label{fig:BPNN2_peeling_angle_max_U}
\end{figure}

\begin{figure}[h!]
	\centering
	\includegraphics[scale=.3]{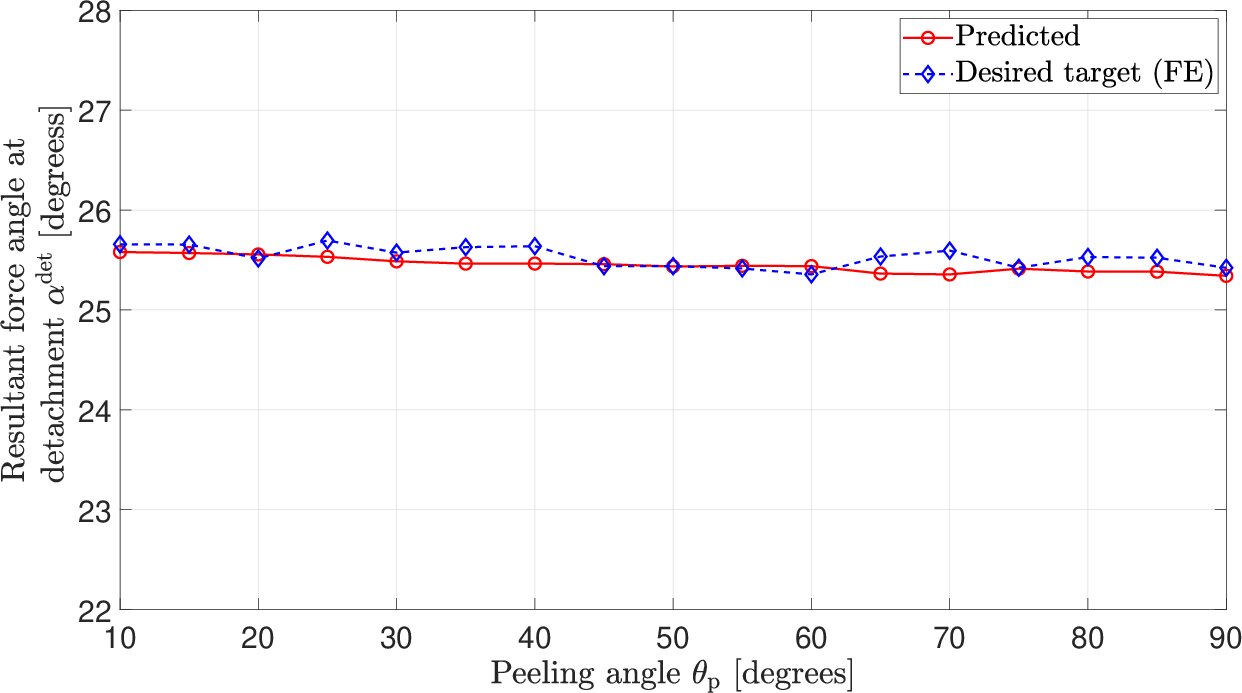}
	\caption{Plot of predicted and desired (FE) value of resultant force angle at detachment  ${\boldsymbol{\alpha }}^{\mathrm{det}}$ with the peeling angle $\theta_\mathrm{p}$ across all the splits for model BR-BPNN-II.}
	\label{fig:BPNN2_peeling_angle_at_detachement}
\end{figure}

Figures~\ref{fig:BPNN2_peeling_angle_max_U} and~\ref{fig:BPNN2_peeling_angle_at_detachement}  show the predictions for the output dataset of BR-BPNN-II, i.e. the applied displacement at detachment $\bar{u}^\mathrm{det}$ and the resultant force angle at detachment $\alpha^\mathrm{det}$ using the corresponding training parameters from Table~\ref{tab:TP_BRBPNN2}. Again, as mentioned previously,  since in the present work a 5-fold cross validation method is used, the predicted and the desired results across all the splits are shown.
It can be seen from Fig.~\ref{fig:BPNN2_peeling_angle_max_U} that the predicted values of ${\overline{u}}^{\mathrm{det}}$ for all the angles  except for $\theta_\mathrm{p} = 10^\circ$ are very close to the desired target FE results.  The maximum, minimum and the average RE values, given in Table~\ref{tab:APE3}, are estimated to be $9.85\%, 0.15\%$, and $1.24\%$, respectively.
Although the predicted results are evaluated for all the splits, the most  effective results are observed for the third split for the case of  $\theta_\mathrm{p} = 10^\circ$. Similar to the reasoning in Section~\ref{results_discussion_case_I}, the pestilential effects of cumulative MSE values pertaining to the first, second, fourth, and the fifth split cannot be compensated by the MSE value in the third split. This results in discrepancies between the neural network predicted outputs and FE-based desired outputs. However, in case of the $\theta_\mathrm{p} = 20^\circ$ angle, this problem is addressed by mitigating the adverse effects of the training splits (first, second, third, and fifth) with the benefits of the smallest MSE value in the testing split (i.e., the fourth split).
As shown in Fig.~\ref{fig:BPNN2_peeling_angle_at_detachement}, the predicted values of $\alpha^\mathrm{det}$ are also very close to the desired target FE results. The maximum, minimum and the average RE values corresponding to the $\alpha^\mathrm{det}$ predictions are estimated to be $0.66\%, 0.06\%$, and $0.30\%$, respectively.  It can be observed that the predictions are very accurate even outside of the training data set.

\begin{table}
\centering
	\caption{Relative error (RE) for the predictions of model BR-BPNN-III.}\label{tab:APE3}
\begin{tabular}{|p{1.5in}|p{1.1in}|p{1.1in}|} \hline
 & $\alpha^\mathrm{det}$ & $\bar{u}^\mathrm{det}$ \\ \hline
Maximum RE (\%)  & $0.66$ & $9.85$ \\ \hline
Minimum RE (\%) & $0.06$ & $0.15$ \\ \hline
Average RE (\%) & $0.30$ & $1.24$ \\ \hline
\end{tabular}
\end{table}

From all these results, it can be observed that for both the BR-BPNN models, the predictions are very close to the target outputs value except for $\theta_\mathrm{p} = 10^\circ$. Further, for both the BR-BPNN-I and BR-BPNN-II models, the deviations in the predictions are larger for displacements rather than forces. Whereas in case of BR- BPNN-I and BR-BPNN-II, $\bar{u}^\mathrm{max}$, and $\bar{u}^\mathrm{det}$ vary quite abruptly near $\theta_\mathrm{p} = 10^\circ$. This is because for both BR-BPNN-I and BR-BPNN-II , $\bar{u}^\mathrm{max}$ and $\bar{u}^\mathrm{det}$ vary quite abruptly at $\theta_\mathrm{p} = 10^\circ$ (as seen in Figures~\ref{fig:BPNN1_peeling_angle_max_U} and ~\ref{fig:BPNN2_peeling_angle_max_U}) and thus can be considered as outliers.

The important advantage of these ANN models lies in the significant reduction in computational cost. It is observed that the time to train the networks with the data corresponding to all the testing peeling angles of each split for both networks is hardly more than one minute. Similarly, once the network is trained, any number of predictions can be made within minutes. Thus, using FE models in conjunction with ANNs has the potential to significantly reduce the computational time leading to faster analysis once the required data has been obtained. This gives a particularly big advantage when the data is obtained using experiments.

\section{Conclusions}
An artificial neural network model is constructed in the present work to study the peeling behavior of a thin strip such as a gecko spatula. In particular, the variation of the maximum normal and tangential pull-off forces, the corresponding applied displacement, the resultant force angle and  the applied displacement at detachment as a function of the peeling angle are investigated. The input data is obtained from the finite element analysis of Gouravaraju et al.~\cite{Gouravaraju2020a,Gouravaraju2020b}. Bayesian regularization in conjunction with $k$-fold cross validation method is used to form two separate networks. The two networks correspond to (a) the maximum normal and tangential pull-off force and the corresponding applied displacement, and (b) the resultant force angle and  the applied displacement at detachment. The number of hidden neurons in each model are evaluated based on their respective mean square errors. From all the results, the maximum and minimum relative deviations of the predicted values from the FE results are found to be $9.85\%$ and $0.0076\%$ respectively. Based on the results, it can be concluded that the  Bayesian regularization-based backpropagation neural networks can be employed to successfully study peeling problems. The present work successfully shows that utilizing ANN algorithms can significantly reduce the computational time. Further, the proposed neural network models can be extended to predict the influence of various geometrical, material, and environmental factors on  gecko spatula peeling. Another interesting problem that can be investigated using BR-BPNN is the constitutive modeling for the hierarchical structures in the gecko adhesion mechanism.

\section*{Acknowledgments}

The authors gratefully acknowledge the support from SERB, DST, under projects SB/FTP/ ETA-0008/2014 and IMP/2019/000276.

\appendix

\section{Background theory on BPNN with Bayesian regularization}\label{app:brbpnn_theory_initial}
A classical neural network architecture mimics the function of the human brain. The brain neurons and their connections with each other form an equivalence relation with neural network neurons and their associated weight values ($\emph{w}$). In a single layer network with multiple neurons, each element $u_j$ of an input vector is associated with each neuron $i$ with a corresponding weight $\emph{w}_{ij}$. A constant scalar term called bias $b_i$ corresponding to each neuron, which is like a weight, is generally introduced in order to increase the flexibility of the network. This bias $b_i$ is multiplied by a scalar input value (chosen to be 1 here) and is added to the weighted sum $ \sum_{j}\emph{w}_{ij}u_j$ of the vector components $u_j$ to form a net input $n_i$. This net input $n_i$ is then passed to an activation function $f$ (also called transfer function) that produces an output value $a_i$. In general, a neural network consists of two or more layers. Adding a hidden layer of neurons between the input layer and output layer constitutes a multi-layer neural network, also named shallow neural network. The addition of more than one hidden layer in the multi-layer neural network is called a deep neural network.

Traditionally, a BPNN model, a kind of multi-layer neural network, comprises three layers: an input layer, one or more hidden layers, and an output layer, as shown in Fig.~\ref{fig:BPNN}.
\setcounter{figure}{0}
\renewcommand{\thefigure}{A\arabic{figure}}
\begin{figure}[htb]
	\begin{center}
		\includegraphics[scale=0.55]{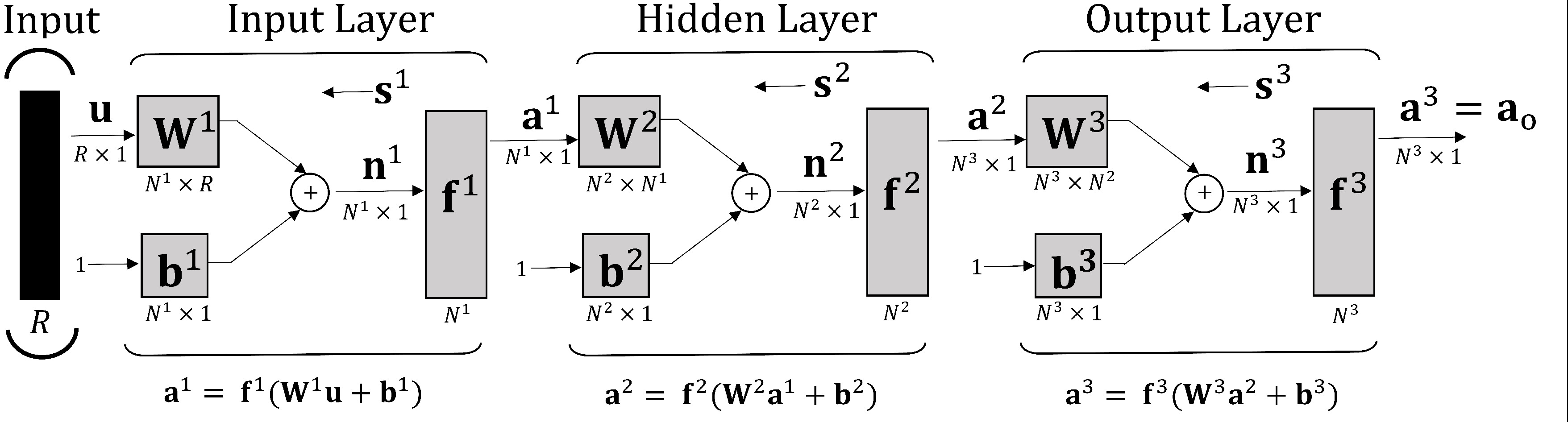}
		\caption{A typical backpropagation neural network with input, hidden, and output layers. Adapted from~\cite{Hagan2014}.}
		\label{fig:BPNN}
	\end{center}
\end{figure}
The input layer associates the input vector $\mathbf{u}$ having $R$ elements with input weight matrix $\mathbf{W}_{}^{1}$ and first bias vector $\mathbf{b}_{}^{1}$ to yield an effective input $\mathbf{n}^1$ to the activation function $\mathbf{f}^1$, which produces an output vector $\mathbf{a}_{}^{1}$. The output vector $\mathbf{a}_{}^{1}$ from the first layer forms the input to the hidden layer and is associated with the weight matrix $\mathbf{W}_{}^{2}$ and bias vector $\mathbf{b}_{}^{2}$ of the hidden layer. At last, the hidden layer output $\mathbf{a}_{}^{2}$ is given as an input to the output layer and delivers a predicted output $\mathbf{a}_{}^{3}$ with weight matrix $\mathbf{W}_{}^{3}$ and bias vector $\mathbf{b}_{}^{3}$. In a neural network with a total of $n_l$ number of layers, the weight matrix $\mathbf{W}^{l}$ and bias vector $\mathbf{b}^l$ for layer $l$ (where $l = 1, 2,\ldots, n_l$) can be written as
\begin{equation}
\mathbf{W}^{l} = \begin{bmatrix}
\mathrm{\emph{w}}_{11}^{\,l} & \mathrm{\emph{w}}^{\,l}_{12} & \mathrm{\emph{w}}^{\,l}_{13} & \ldots & \mathrm{\emph{w}}^{\,l}_{1R} \\[1.4ex]
\mathrm{\emph{w}}^{\,l}_{21} & \mathrm{\emph{w}}^{\,l}_{22} & \mathrm{\emph{w}}^{\,l}_{23} & \ldots &\mathrm{\emph{w}}^{\,l}_{2R} \\
\vdots & \vdots & \vdots & \ddots & \vdots\\[1.4ex]
\mathrm{\emph{w}}\!_{N^{l} 1}^{\,\,l} & \mathrm{\emph{w}}\!_{N^{l} 2}^{\,\,l} & \mathrm{\emph{w}}\!_{N^{l} 3}^{\,\,l} & \ldots & \mathrm{\emph{w}}\!_{N^{l} R}^{\,\,l}
\end{bmatrix}\,,
\quad \quad
\mathbf{b}^{\,l} = \begin{bmatrix}
b_1^{\,l} \\[1.4ex]
b_2^{\,l} \\[1.4ex]
\vdots \\[1.4ex]
b\!_{N^{l}}^{\,\,l}
\end{bmatrix}\,,
\end{equation}
where $N^{l}$ denotes the number of neurons in layer ${l}$ and the effective input $\mathbf{n}^{l}$ is then given as
\begin{equation}
\mathbf{n}^{l} = \mathbf{W}^{l}\mathbf{a}^{l-1} + \mathbf{b}^{l}\,, \quad \quad \mathrm{with}~~ \mathbf{a}^0 = \mathbf{u}\,.
\end{equation}

The number of neurons in the input layer ($N^1$) and output layer ($N^3$) is linked to the number of input and output vectors, respectively. However, the number of neurons in the hidden layer ($N^2$) are accountable for the quantification of the weights and biases. The optimal network structure is versed by the optimum number of neurons in each layer required for the training and denoted as $N^1$-$N^2$-$N^3$. A variety of activation functions are used in backpropagation neural network, viz., hard limit, linear, sigmoid, log-sigmoid, hyperbolic tangent sigmoid~\cite{Hagan2014}. In the current work, linear activation functions are employed in all the layers according to which, the output is equal to the input i.e. $\mathbf{a}^{l} = \mathbf{n}^{l}$.

The network error $\mathbf{e}$ is calculated by subtracting predicted output $\mathbf{a}_\mathrm{o}$ from target output $\mathbf{t}_\mathrm{o}$. The sensitivity $\mathbf{s}$, which measures how the output of the network changes due to perturbations in the input, is back-propagated from output layer ($\mathbf{s}^3$) to input layer ($\mathbf{s}^1$) via the hidden layer ($\mathbf{s}^2$). Through the backpropagation process, the error of the neurons in the hidden layer is estimated as the backward weighted sum of the sensitivity. Thereafter, to update weights, different learning algorithms are used in association with the sensitivity such as the steepest descent, LM, and conjugate gradient algorithms. The sensitivity at layer $l$ is calculated using the recurrence relation~\cite{Hagan2014}
\begin{eqnarray}
\mathbf{s}^{l} & =& \mathbf{\dot{F}}^{l}\!\left(\mathbf{n}^{l}\right)\, \mathbf{W}^{l+1}\, \mathbf{s}^{l+1}\,, \quad \mathrm{where}\,\, l = n_l-1, \ldots ,2,1 \,,\label{eq:sensitivities1} \\
\mathrm{with} \quad \mathbf{s}^{n_l} & =& \mathbf{\dot{F}}^{n_l}\big(\mathbf{n}^{n_l}\big)\, \left(\mathbf{t}_\mathrm{o} - \mathbf{a}_\mathrm{o}\right)\,,
\label{eq:sensitivities2}
\end{eqnarray}
where $\mathbf{\dot{F}}^{l}(\mathbf{n}^{l})$ is a diagonal matrix containing the partial derivatives of the activation function $\mathbf{f}^{l}$ with respect to the net inputs $\mathbf{n}^{l}$ and is given as
\begin{equation}
\mathbf{\dot{F}}^{l}\big(\mathbf{n}^{l}\big) = \begin{bmatrix}
\dot{f}^{l}\!\left(n_1^{l}\right) & 0 & \ldots & 0\\
0 & \dot{f}^{l}\!\left(n_2^{l}\right) & \ldots & 0\\
\vdots & \vdots & \ddots & \vdots\\
0 & 0 & \ldots & \dot{f}^{l}\!\left(n\!_{N^{l}}^{\,\,l}\right)
\end{bmatrix}\,,
\quad \quad \mathrm{where} \quad \dot{f}^{l}\!\left(n_j^{l}\right) = \frac{\partial f^l\!\left(n_j^{l}\right)}{\partial n_j^{l}}\,,
\end{equation}
and for the considered linear activation function is equal to the identity matrix.

The purpose of a backpropagation neural network model is to ensure a network with small deviations for the training dataset and supervise the unknown inputs effectively. The intricacy of the BPNN, monitored by neurons in the hidden layer and their associated weights, leads to overfitting, i.e. the network tries to make the error as small as possible for the training set but performs poorly when new data is presented. However, a robust network model should be able to generalize well, i.e. it should predict well even when presented with new data. Therefore, Bayesian regularization based learning of BPNN models is utilized to achieve better generalization and minimal over-fitting for the trained networks~\cite{MacKay1992,Burden2008}.

\section{Results from finite element simulations}\label{app:A}
Table~\ref{tab:FEdata} lists the values of the maximum normal force $F_\mathrm{n}^\mathrm{max}$, maximum tangential force $F_\mathrm{t}^\mathrm{max}$, applied displacement at force maximum $\bar{u}^\mathrm{max}$, applied displacement at $\bar{u}^\mathrm{det}$, and resultant force angle at detachment $\alpha^\mathrm{det}$ for different peeling angles as obtained by Gouravaraju et al.~\cite{Gouravaraju2020a,Gouravaraju2020b} using nonlinear finite element analysis.
\setcounter{table}{0}
\renewcommand{\thetable}{B\arabic{table}}
\fontsize{9}{11}\selectfont{
\begin{longtable}[pos=h]{c c c c c c c}
		\caption{Data from finite element results of Gouravaraju et al.~\cite{Gouravaraju2020a,Gouravaraju2020b}.\label{tab:FEdata}} \\
		\hline
		 & Peeling  & Applied  & Max. normal & Max. tangential & Applied   & Resultant force \\
		Case & angle & displacement at  & pull-off force & pull-off force  & displacement at  & angle at detachment  \\
		& $\theta_\mathrm{p}$ & force max.  & $F_\mathrm{n}^\mathrm{max}$ &$F_\mathrm{t}^\mathrm{max}$ & at detachment & $\alpha^\mathrm{det}$\\
		& $\mathrm{[degrees]}$& $\bar{u}^\mathrm{max}$ [nm] & [nN] & [nN] & $\bar{u}^\mathrm{det}$ [nm] &  [degrees]\\	\hline	
		$1$ & $10$ & $41.8$ & $174.1584$ & $1722.719$ & $393.4$ & $25.64973$ \\
		$2$ & $15$ & $35.6$ & $171.1613$ & $1529.699$ & $263.8$ & $25.57726$ \\
		$3$ & $20$ & $33.8$ & $165.5169$ & $1370.545$ & $199.6$ & $25.56427$ \\
		$4$ & $25$ & $32.4$ & $160.1255$ & $1240.153$ & $161.6$ & $25.59890$ \\
		$5$ & $30$ & $31.2$ & $155.0284$ & $1129.391$ & $136.6$ & $25.60988$ \\
		$6$ & $35$ & $30.6$ & $150.3356$ & $1034.944$ & $119.0$ & $25.55115$ \\
		$7$ & $40$ & $30.2$ & $145.7655$ & $950.3074$ & $106.2$ & $25.55958$ \\
		$8$ & $45$ & $30.0$ & $141.2537$ & $872.9422$ & $96.6$  & $25.61840$ \\
		$9$ & $50$ & $30.2$ & $136.8172$ & $801.4117$ & $89.2$  & $25.65779$ \\
		$10$ & $55$ & $30.6$ & $132.2554$ & $733.2346$ & $83.4$  & $25.62680$ \\
		$11$ & $60$ & $31.4$ & $127.5051$ & $667.3803$ & $78.8$  & $25.53845$ \\
		$12$ & $65$ & $32.8$ & $122.5176$ & $602.7001$ & $75.4$  & $25.66338$ \\
		$13$ & $70$ & $34.4$ & $117.0706$ & $537.6730$ & $72.6$  & $25.51802$ \\
		$14$ & $75$ & $36.8$ & $110.9777$ & $471.2534$ & $70.6$  & $25.49363$ \\
		$15$ & $80$ & $40.2$ & $104.0514$ & $402.4569$ & $69.4$  & $25.69447$ \\
		$16$ & $85$ & $44.8$ & $95.87533$ & $330.1474$ & $68.4$  & $25.44845$ \\
		$17$ & $90$ & $51.8$ & $86.18540$ & $254.5306$ & $68.2$  & $25.49894$ \\
		\hline
\end{longtable}
}

\fontsize{12}{12}\selectfont

\section{Framework of Bayesian regularization-based backpropagation}\label{app:B}
The algorithm for the Bayesian regularization based backpropagation is composed of the following steps:
\begin{enumerate}
	\item Pick training data set $D$ containing the $N_{\mathrm{train}}$ cases specified in Tables~\ref{tab:kfold_details}, \ref{tab:out_data_1} and \ref{tab:out_data_2}, and Appendix~\ref{app:A}.
	\begin{itemize}
		\item[(a)] Input vector, $\mathbf{u}$: Peeling angles $\theta_\mathrm{p}$
		\item[(b)] Target output vector, $\mathbf{t}_\mathrm{o}$ \,: $\bar{\boldsymbol{u}}^\mathrm{max}$, $\boldsymbol{F}_{\mathrm{n}}^\mathrm{max}$, $\boldsymbol{F}_{\mathrm{t}}^\mathrm{max}$ (for BR-BPNN-I)
		\item[] \hspace{4.8cm} $\bar{\boldsymbol{u}}^\mathrm{det}$\,, $\boldsymbol{\alpha}^\mathrm{det}$ (for BR-BPNN-II)
	\end{itemize}

	\item Initialize neural network with
	 \begin{itemize}
	 	\item[(a)] Number of neurons in the input layer equal to the number of input vectors, which is equal to 1 for both the BR-BPNN models as described in step 1(a), i.e. $N^1 = 1$.
	 	\item[(b)] Number of neurons in the output layer equal to the number of output vectors, which is equal to $N^3 = 3$  for model BR-BPNN I and $N^3 = 2$ for model BR-BPNN II,  respectively, as described in Tables~\ref{tab:out_data_1} and~\ref{tab:out_data_2}.
	 	\item[(c)] Number of neurons in the hidden layer equal to one, i.e. $N^2 = 1$.
	 \end{itemize}
	\item Set learning method to Bayesian regularization
	\begin{itemize}
		\item[(a)] Set maximum number of epochs to 2000.
		\item[(b)] Divide the training data set as per Table~\ref{tab:kfold_details} using $k$-fold cross validation.
	\end{itemize}
	\item Train the network
	\begin{itemize}
		\item[(a)] Compute regularization parameters $\mu$ and $\nu$ using Eq.~(\ref{eq:reg_param}).
		\item[(b)] Backpropagate sensitivities calculated using Eqs.~(\ref{eq:sensitivities1}) and (\ref{eq:sensitivities2}).
		\item[(c)] Update weights using Eq.~(\ref{eq:weight_update}).
	\end{itemize}
	\item Compute mean square error (MSE) using Eq.~(\ref{eq:MSE}).
	\item Loop over steps 4 and 5 with different number of neurons in the hidden layer.
	\item Plot the MSE with number of neurons in the hidden layer as in Fig.~\ref{fig:MSE}.
	\item Select the number of neurons in the hidden layer to be the value from which MSE attains a broad minimum and decreases as $N^2$ is further increased. This determines the optimal network structure $N^1$-$N^2$-$N^3$.
	\item Retrain the neural network model with optimal network structure from step 8.
	\item Save the model parameters (using Tables~\ref{tab:TP_BRBPNN1} and~\ref{tab:TP_BRBPNN2}) along with weights and biases.
	\item Using the saved parameters in step 10, predict for the testing dataset as in Tables~\ref{tab:kfold_details}, ~\ref{tab:out_data_1} and~\ref{tab:out_data_2}.
\end{enumerate}

%
%
%
%
%
%
%
%
%
%


\bibliographystyle{SachinJA2021}
\bibliography{MLpaper}

\addcontentsline{toc}{chapter}{References}


\end{document}